\documentclass[article]{JHEP3}

\usepackage{amsmath,amssymb}
\usepackage[dvips]{graphics}
\usepackage{epsfig}
\usepackage{graphicx}% Include figure files
\usepackage{bm}
\newcommand{\be}{\begin{equation}}
\newcommand{\ee}{\end{equation}}
\newcommand{\bea}{\begin{eqnarray}}
\newcommand{\eea}{\end{eqnarray}}
\newcommand{\bean}{\begin{eqnarray*}}
\newcommand{\eean}{\end{eqnarray*}}

\def\beq{\begin{equation}}

\def\eeq{\end{equation}}

\relax

\def\be{\begin{equation}}
\def\bearl{\begin{array}{l}}
\def\bearll{\begin{array}{ll}}
\def\ee{\end{equation}}
\def\eear{\end{array}}

\preprint{{\small DF/IST-17.2003}\\ 
{\small \texttt{hep-th/0311260}}}

\title{Scalar-gravitational perturbations 
and quasinormal modes in the five dimensional 
Schwarzschild black hole}

\author{Vitor Cardoso$^{\dag}$, Jos\'e P. S. Lemos$^{\dag}$ 
and Shijun Yoshida$^{\dag}$ 
\\
$^{\dag}$Centro Multidisciplinar de Astrof\'{\i}sica - CENTRA, 
Departamento de F\'{\i}sica, Instituto Superior T\'ecnico,\\ 
Av. Rovisco Pais 1, 1096 Lisboa, Portugal\\
\\
\email{vcardoso@fisica.ist.utl.pt}, \quad
\email{lemos@kelvin.ist.utl.pt}, \quad 
\email{yoshida@fisica.ist.utl.pt}
}

\abstract{We calculate the quasinormal modes (QNMs) for gravitational
perturbations of the Schwarzschild black hole in the five dimensional
(5D) spacetime with a continued fraction method. 
For all the types of perturbations (scalar-gravitational, 
vector-gravitational, and tensor-gravitational perturbations), the
QNMs associated with $l=2$, $l=3$, and $l=4$ are
calculated. Our numerical results are summarized as follows: (i) The
three types of gravitational perturbations associated with the same
angular quantum number $l$ have a different set of the 
quasinormal (QN) frequencies; (ii) There is no purely imaginary frequency
mode; (iii) The three types of gravitational perturbations have the
same asymptotic behavior of the QNMs in the limit of the
large imaginary frequencies, which are given by $\omega
T_H^{-1}\rightarrow\log{3}+\,2\pi i (n+1/2)$ as $n\rightarrow\infty$,
where $\omega$, $T_H$, and $n$ are the oscillation frequency, the
Hawking temperature of the black hole, and the mode number,
respectively.
}


\begin{document}

\vfill

\eject

\section{Introduction}

The quasinormal (QN) ringing of a Schwarzschild spacetime was first
observed in \cite{vish} through numerical calculations of
the gravitational wave scattering by the black hole. Since then, the
quasinormal modes (QNMs) of black holes have been extensively
studied. The classical motivation behind the exploration of the QNMs
of black holes is twofold: One is to answer the question of whether
the spacetime is stable, and the other to know what kind of
oscillations will be excited in the spacetime as some perturbations
are given. The latter is quite important from the observational point
of view because we could determine fundamental parameters of a black
hole, such as the mass or the angular momentum, through the
information of the QNMs. A number of studies on QNMs of
several different spacetimes containing black holes have been 
done. (for a review see, e.g.,  \cite{kokkotas99,nollert99}).  
Recently QNMs have acquired a different status, since it was 
conjectured that they may be connected to black hole area quantization 
and quantum gravity \cite{hod,CLY03,All}.

Most studies on QNMs of black holes were restricted to the four
dimensional (4D) case, compatible with astrophysical
scenarios. However, motivated by the
TeV-scale gravity proposals \cite{hamed} for instance, 
higher-dimensional (higher-D) theories of gravity have
recently attracted much attention.  Up to recently, there were
no master equations for examining the QNMs for gravitational
perturbations of higher-D black holes.  Due to the absence
of these master equations, only the QNMs of the simplest test fields,
namely massless scalar fields, around the higher-D black
holes were calculated \cite{CDL03,IUM03,Ko03a,MN03}.

The situation has changed recently. Kodama and Ishibashi \cite{KI03a}
have derived master equations for gravitational perturbations in a
higher-D Schwarzschild black hole spacetime.  Putting $D=2+n$, it was
shown that for the case $n\ge3$, the gravitational perturbations
can be divided into three classes, namely, scalar-gravitational,
vector-gravitational, and tensor-gravitational perturbations,
according to their tensorial behavior on the $n$-sphere. The
scalar-gravitational and vector-gravitational perturbations correspond
to the polar and axial perturbations in the 4D spacetime. The
tensor-gravitational perturbations are a new kind appearing in
higher-Ds.  Kodama and Ishibashi have also shown that all the types of
gravitational perturbations can be reduced to a simple
Schr\"odinger-type wave equation like Regge-Wheeler or Zerilli
equations. It is important to ask the question of whether there is a
special relationship among the scalar-gravitational,
vector-gravitational, and tensor-gravitational perturbations because
in 4D the scalar-gravitational and vector-gravitational perturbations
have a special relation yielding the same quasinormal (QN)
frequencies. There seems to be no such relation between the
scalar-gravitational and vector-gravitational perturbations
\cite{KI03a}.

Recently, some studies on the QNMs for gravitational
perturbations of higher-D Schwarzschild black holes have
appeared. Konoplya has calculated the fundamental QNMs of Kodama and
Ishibashi's master equations with a JWKB approximation, and has found
that three types of perturbations have different fundamental
QN frequencies \cite{Ko03b}.  Berti, Cavagli\`a, and Gualtieri have also 
done similar calculations but for a wide range of angular eigenvalues $l$ 
of perturbations \cite{BCG03}. Cardoso, Lemos, and Yoshida have
calculated the QNMs for the vector-gravitational and
tensor-gravitational perturbations up to higher-order modes
\cite{CLY03}.  As for the asymptotic behavior in the limit of highly
damped modes, it has been shown that the three types of gravitational
perturbations have the same behavior regardless of the spacetime
dimensions or the angular quantum numbers of the perturbations
\cite{Bi03,CLY03}.

The purpose of this paper is to go a step further up and explore the
QNMs in the special 5D higher-D black hole spacetime. In
particular, we want to find out whether or not there is a relationship
among the QNMs of the scalar-gravitational, vector-gravitational and
tensor-gravitational perturbations in this 5D Schwarzschild black
hole.  The 5D case can be considered representative of the other
higher-D cases.  A prime purpose of the present study is to
extensively calculate the QNMs for the scalar-gravitational
perturbations. The scalar-gravitational perturbations are the most 
important because they can be excited more easily than other types of 
perturbations. 
We calculate the QN frequencies not only for low-order
modes but also for relatively higher-order modes, enabling us to
discuss the asymptotic behavior of the highly damped QNMs. In the
present study Nollert's approach is employed numerically to obtain
the QN frequencies.  The paper is organized as follows: In section 2
we present the basic equations employed for obtaining QNMs in the 5D
Schwarzschild spacetime. Numerical results are given in section 3, and
in section 4 we conclude.

%%%%%%%%%%%%%%%%%%%%%%%%%%%%%%%%%%%%%%%%%%%%%%%%%%%%%
\section{Numerical Method}
%%%%%%%%%%%%%%%%%%%%%%%%%%%%%%%%%%%%%%%%%%%%%%%%%%%%%%
The general perturbation equations for the Schwarzschild black hole in
$D=n+2$ dimensions have been recently derived by Kodama and Ishibashi
\cite{KI03a}. It has been shown that there are three completely
decoupled classes in the perturbations for the $n\geq3$ case, namely
scalar-gravitational, vector-gravitational, and tensor-gravitational
perturbations, and that all types of perturbations can be cast into a
Schr\"odinger-type wave equation, given by
\begin{equation}
f{d\over dr}\left(f{d\Phi\over dr}\right)+(\omega^2-V(r))\Phi=0\,,  
\label{master-eq} 
\end{equation}
where the function $f$ is minus the $g_{tt}$ component of the metric tensor 
in Schwarzschild coordinates, 
and given by  
\begin{equation}
f=1-{2M\over r^{n-1}}=1-x\,. 
\end{equation}
Here, $r$ is the radial coordinate of the $n+2$ dimensional Schwarzschild 
spacetime, and $\omega$ is the oscillation frequency of the perturbations. 
$M$ is a parameter related to the mass $\tilde{M}$ of the black hole
via the relation
\begin{equation}
\tilde{M}={nMA_n\over 8\pi G_{n+2}}\,, 
\end{equation}
where $A_n$ is the area of a unit $n$-sphere, given by 
$A_n=2\pi^{(n+1)/2}/\Gamma[(n+1)/2]$, and $G_{n+2}$ stands for 
the $n+2$ dimensional Newton constant. The Hawking temperature of the black 
hole is $T_H=(n-1)/4\pi r_h$, where $r_h$ is the horizon radius, defined by 
$r_h^{n-1}=2 M$.

In the master equation (\ref{master-eq}), depending on the type of 
perturbations, the effective potential can be written as 
\begin{equation}
{r^2V\over f}=\left\{
\begin{array}{ll}
{Q(r)\over 16 H(r)^2}&{\rm for\ scalar} \\
{(2l+n)(2l+n-2)\over 4}-{3n^2x\over 2}&{\rm for\ vector} \\
{(2l+n)(2l+n-2)\over 4}+{ n^2x\over 2}&{\rm for\ tensor}\,,
\end{array}
\right.
\label{def-potential}
\end{equation}
with 
\begin{eqnarray}
Q(r)&=&n^4(n+1)^2x^3+n(n+1)\{4(2n^2-3n+4)m+n(n-2)(n-4)(n+1)\}x^2 \\ \nonumber
&-&12n\{(n-4)m+n(n+1)(n-2)\}mx+16m^3+4n(n+2)m^2\,,  \\
H(r)&=&m+{1\over 2}n(n+1)x\,,\\
m&=&l(l+n-1)-n\,, 
\end{eqnarray}
where $l$ means the angular quantum number of perturbations, and $l\ge
2$ has been assumed. It is worthwhile to note that the 
scalar-gravitational and vector-gravitational perturbations are,
respectively, higher-D counterparts of the polar and axial
perturbations in the 4D Schwarzschild black hole, while
the tensor-gravitational perturbations are brand new in the sense that
a counterpart of the tensor-gravitational perturbations does not
appear in the 4D case. In the case of $n=2$, the master
equations for the scalar-gravitational and vector-gravitational
perturbations therefore become the Zerilli equation and the 
Regge-Wheeler equation,
respectively \cite{Ze70,RW57}. Interestingly, the effective potential
for the tensor-gravitational perturbations is exactly the same as that for the
massless scalar field in the higher-D Schwarzschild black
hole \cite{CDL03}.

The QNMs of the higher-D Schwarzschild black hole are characterized 
by the boundary conditions of incoming waves at the black hole horizon and 
outgoing waves at spatial infinity, written as
\begin{equation}
\Phi(r)\rightarrow\left\{
\begin{array}{ll}
e^{-i\omega r_*}&{\rm as}\ r_*\rightarrow \infty\\
e^{ i\omega r_*}&{\rm as}\ r_*\rightarrow-\infty\,, 
\end{array}
\right.
\label{def-bc}
\end{equation}
where the time dependence of perturbations has been assumed to be 
$e^{i\omega t}$. Here, $r_*$ denotes the tortoise coordinate, defined 
by $dr_*=f^{-1}dr$. 
In order to obtain numerically the QN frequencies, 
we employ Nollert's method 
\cite{No93}, an enhanced version of Leaver's continued fraction 
method \cite{Le85}, since it is nicely suitable to the study of 
the asymptotic behavior of QNMs in the limit of highly 
damped modes, the main concern in the present investigation.

In this study, we only consider the QNMs of the Schwarzschild black 
hole in the five dimensional spacetime, namely the $n=3$ case. The tortoise 
coordinate is then reduced to 
\begin{equation}
r_*=z^{-1}+{1\over 2z_1}\ln(z-z_1)-{1\over 2z_1}\ln(z+z_1)\,, 
\end{equation}
where $z=r^{-1}$ and $z_1=r_h^{-1}$. The perturbation function $\Phi$ 
can be expanded around the horizon as 
\begin{equation}
\Phi=e^{-i\omega z^{-1}}(z-z_1)^\rho(z+z_1)^\rho\sum_{k=0}^\infty 
a_k\left({z-z_1\over-z_1}\right)^k\,, 
\label{expansion}
\end{equation}
where $\rho=i\omega/2z_1$ and $a_0$ is taken to be $a_0=1$. Here, the
expansion coefficients $a_k$ for $k\ge 1$ are determined with the
recurrence relation, the order of which is dependent on the functional
form of the effective potential in equation (\ref{master-eq}).  As
shown by Cardoso, Lemos and Yoshida \cite{CLY03}, the recurrence
relation becomes a four-term relation for vector-gravitational and
tensor-gravitational perturbations. In this paper we do not show an
explicit expression of the four-term recurrence relation for the
vector-gravitational and tensor-gravitational perturbations since it
can be found in \cite{CLY03}. For the scalar-gravitational
perturbations, on the other hand, the recurrence relation becomes
eight-term relation, given by 
\begin{eqnarray}
&&c_0(0)a_1+c_1(0)a_0=0\,, \nonumber \\
&&c_0(1)a_2+c_1(1)a_1+c_2(1)a_0=0\,, \nonumber \\
&&c_0(2)a_3+c_1(2)a_2+c_2(2)a_1+c_3(2)a_0=0\,, \nonumber \\
&&c_0(3)a_4+c_1(3)a_3+c_2(3)a_2+c_3(3)a_1+c_4(3)a_0=0\,, \nonumber \\
&&c_0(4)a_5+c_1(4)a_4+c_2(4)a_3+c_3(4)a_2+c_4(4)a_1+c_5(4)a_0=0\,, \nonumber \\
&&c_0(5)a_6+c_1(5)a_5+c_2(5)a_4+c_3(5)a_3+c_4(5)a_2+c_5(5)a_1+c_6(5)a_0=0\,, \nonumber \\
&&c_0(k)a_{k+1}+c_1(k)a_k+c_2(k)a_{k-1}+c_3(k)a_{k-2}+c_4(k)a_{k-3}+c_5(k)a_{k-4}\, \nonumber \\
&&+c_6(k)a_{k-5}+c_7(k)a_{k-6}=0\,\,,{\rm for\ } k=6,7\cdots, 
\label{recu}
\end{eqnarray}

where 
\begin{eqnarray}
c_0(k)&=&2(1+k)(6+m)^2(1+k+2\rho)\,, \nonumber \\
c_1(k)&=&-(6+m)\{m^2+(78+5m)k^2+10m(1+2\rho)\rho+(30+5m+216\rho+20m\rho)k+
12(1+5\rho+10\rho^2)\}\,, \nonumber \\
c_2(k)&=&[8(324+48m+m^2)k^2+m^2(1+32\rho^2)+12m(13-48\rho+112\rho^2)+36(41-128\rho+192\rho^2)
\nonumber \\
&&+16\{-198+540\rho+2m^2\rho+3m(-7+30\rho)\}k]/2\,, \nonumber \\
c_3(k)&=&[-4(1980+168m+m^2)(-2+k)^2-m^2(3+4\rho)^2-12m(69+192\rho+224\rho^2) \nonumber \\
&&-36(129+520\rho+752\rho^2)-4(-2+k)\{m^2(3+4\rho)+96m(3+7\rho)
+36(65+204\rho)\} ]/4 \,, \nonumber \\
c_4(k)&=&12[6(25+m)k^2+m(25-44\rho+24\rho^2)+6(127-214\rho+96\rho^2)\nonumber \\
&&+\{-660+588\rho+m(-22+24\rho)\}k] \,, \nonumber \\
c_5(k)&=&-3[4(81+m)k^2+m(29-40\rho+16\rho^2)+6(491-644\rho+216\rho^2)\nonumber \\
&&+4\{-483+324\rho+m(-5+4\rho)\}k ]\,, \nonumber \\
c_6(k)&=&18\{227+16k^2-240\rho+64\rho^2+8(-15+8\rho)k\}\,, \nonumber \\
c_7(k)&=&-9(-9+2k+4\rho)^2\,. \nonumber \\
\end{eqnarray}
It is understood that since the asymptotic form of the perturbations as 
$r_*\rightarrow\infty$ is written in terms of the variable $z$ as
\begin{equation}
e^{-i\omega r_*}=e^{-i\omega z^{-1}}(z-z_1)^{-\rho}(z+z_1)^{\rho}\,,
\label{expwr} 
\end{equation}
the expanded perturbation function $\Phi$ defined by equation 
(\ref{expansion}) automatically satisfy the QNM boundary conditions 
(\ref{def-bc}) if the power series converges for $0\le z\le z_1$.
After making a Gaussian elimination five times \cite{Le90}, we can 
reduce the eight-term recurrence relation to a three-term relation, which 
has the form 
\begin{eqnarray}
&&\alpha_0a_1+\beta_0a_0=0\,, \nonumber \\
&&\alpha_ka_{k+1}+\beta_ka_k+\gamma_ka_{k-1}=0\,, 
\ k=1,2,\cdots.  
\end{eqnarray}
Here, we omit the explicit expression for the final three-term 
recurrence relation because the numerical procedure to obtain the 
three-term recurrence relation is quite simple. 
Now that we have the three-term recurrence relation for determining the 
expansion coefficients $a_k$, according to Leaver \cite{Gu67,Le85}, the 
convergence condition for the expansion (\ref{expansion}), namely the 
QNM conditions, can be written in terms of the continued fraction as 
\begin{eqnarray}
0=\beta_0-{\alpha_0\gamma_1\over\beta_1-}
{\alpha_1\gamma_2\over\beta_2-}{\alpha_2\gamma_3\over\beta_3-}
\cdots \,.  
\label{a-eq}
\end{eqnarray}
In order to use Nollert's method, with which relatively higher-order 
QNMs with large imaginary frequencies can be obtained, we have to know 
the asymptotic behavior of $a_{k+1}/a_k$ in the limit of 
$k\rightarrow\infty$. Following Leaver \cite{Le90}, it is found that 
\begin{eqnarray}
{a_{k+1}\over a_k}=1\pm2\sqrt{\rho}\,k^{-1/2}+
\left(2\rho-{3\over 4}\right)k^{-1}+\cdots\,, 
\label{exp-R}
\end{eqnarray}
where the sign for the second term in the right-hand side is chosen so as 
to be 
\begin{eqnarray}
{\rm Re}(\pm2\sqrt{\rho}) < 0\,. 
\end{eqnarray}

In actual numerical computations, it is convenient to solve the $k$-th 
inversion of the continue fraction equation (\ref{a-eq}), given by 
\begin{eqnarray}
&&\beta_k-{\alpha_{k-1}\gamma_k\over\beta_{k-1}-}
{\alpha_{k-2}\gamma_{k-1}\over\beta_{k-2}-}\cdots
{\alpha_0\gamma_1\over\beta_0} \nonumber \\
&=& {\alpha_{k}\gamma_{k+1}\over\beta_{k+1}-}
{\alpha_{k+1}\gamma_{k+2}\over\beta_{k+2}-}\cdots\,. \quad 
\label{a-eq2}
\end{eqnarray}
The asymptotic form (\ref{exp-R}) plays an important role in Nollert's 
method when the infinite continued fraction in the right-hand side of
equation (\ref{a-eq2}) is evaluated \cite{No93}. 

%%%%%%%%%%%%%%%%%%%%%%%%%%%%%%%%%%%%%%%%
\section{Numerical Results}
%%%%%%%%%%%%%%%%%%%%%%%%%%%%%%%%%%%%%%%
\begin{table}
\centering
\caption{\label{tab:scalar}QN frequencies $\omega r_h$ of 
the scalar-gravitational
perturbations for $l=2$, $l=3$ and $l=4$.}
%\begin{ruledtabular}
\begin{tabular}{lccc}
$n$ & $l=2$ & $l=3$ & $l=4$ \\ \hline
$ 0$& $ 0.9477+0.2561i$ & $ 1.6056+0.3110i$ & $ 2.1924+0.3293i$ \\
$ 1$& $ 0.8512+0.8212i$ & $ 1.5109+0.9528i$ & $ 2.1149+0.9999i$ \\
$ 2$& $ 0.6727+1.5431i$ & $ 1.3325+1.6586i$ & $ 1.9649+1.7078i$ \\
$ 3$& $ 0.5121+2.4399i$ & $ 1.1107+2.4693i$ & $ 1.7593+2.4804i$ \\
$ 4$& $ 0.4169+3.4140i$ & $ 0.9091+3.3823i$ & $ 1.5319+3.3352i$ \\
$ 5$& $ 0.3619+4.4092i$ & $ 0.7586+4.3533i$ & $ 1.3215+4.2642i$ \\
$ 6$& $ 0.3275+5.4102i$ & $ 0.6524+5.3471i$ & $ 1.1486+5.2403i$ \\
$ 7$& $ 0.3043+6.4130i$ & $ 0.5765+6.3485i$ & $ 1.0130+6.2392i$ \\
$ 8$& $ 0.2876+7.4164i$ & $ 0.5204+7.3524i$ & $ 0.9071+7.2474i$ \\
$ 9$& $ 0.2751+8.4197i$ & $ 0.4778+8.3569i$ & $ 0.8232+8.2584i$ \\
$10$& $ 0.2653+9.4229i$ & $ 0.4445+9.3614i$ & $ 0.7555+9.2698i$ \\
$11$& $ 0.2574+10.426i$ & $ 0.4178+10.366i$ & $ 0.7000+10.281i$ \\
$12$& $ 0.2510+11.429i$ & $ 0.3961+11.370i$ & $ 0.6537+11.290i$ \\
\end{tabular}
%\end{ruledtabular}
\end{table}

\begin{table}
\centering
\caption{\label{tab:vector}QN frequencies $\omega r_h$ of 
the vector-gravitational
perturbations for $l=2$, $l=3$ and $l=4$.}
%\begin{ruledtabular}
\begin{tabular}{lccc}
$n$ & $l=2$ & $l=3$ & $l=4$ \\ \hline
$ 0$& $ 1.1340+0.3275i$ & $ 1.7254+0.3338i$ & $ 2.2805+0.3400i$ \\
$ 1$& $ 0.9474+1.0220i$ & $ 1.6181+1.0199i$ & $ 2.2003+1.0317i$ \\
$ 2$& $ 0.5429+1.9247i$ & $ 1.4136+1.7655i$ & $ 2.0452+1.7598i$ \\
$ 3$& $ 0.4357+3.1209i$ & $ 1.1485+2.6098i$ & $ 1.8314+2.5508i$ \\
$ 4$& $ 0.3996+4.1798i$ & $ 0.8798+3.5543i$ & $ 1.5914+3.4220i$ \\
$ 5$& $ 0.3996+4.1798i$ & $ 0.6275+4.5609i$ & $ 1.3630+4.3662i$ \\
$ 6$& $ 0.3715+5.2168i$ & $ 0.3095+5.6127i$ & $ 1.1668+5.3579i$ \\
$ 7$& $ 0.3502+6.2441i$ & $ 0.3160+7.0116i$ & $ 1.0034+6.3729i$ \\
$ 8$& $ 0.3335+7.2654i$ & $ 0.3597+8.0670i$ & $ 0.8651+7.3974i$ \\
$ 9$& $ 0.3202+8.2827i$ & $ 0.3696+9.1021i$ & $ 0.7430+8.4249i$ \\
$10$& $ 0.3093+9.2970i$ & $ 0.3695+10.129i$ & $ 0.6290+9.4526i$ \\
$11$& $ 0.3002+10.309i$ & $ 0.3657+11.151i$ & $ 0.5140+10.480i$ \\
$12$& $ 0.2925+11.320i$ & $ 0.3604+12.169i$ & $ 0.3821+11.508i$ \\
\end{tabular}
%\end{ruledtabular}
\end{table}

\begin{table}
\centering
\caption{\label{tab:tensor}QN frequencies $\omega r_h$ of the
tensor-gravitational perturbations for $l=2$, $l=3$ and $l=4$.}
%\begin{ruledtabular}
\begin{tabular}{lccc}
$n$ & $l=2$ & $l=3$ & $l=4$ \\ \hline
$ 0$& $ 1.5106+0.3575i$ & $ 2.0079+0.3558i$ & $ 2.5063+0.3550i$ \\
$ 1$& $ 1.3927+1.1046i$ & $ 1.9170+1.0853i$ & $ 2.4327+1.0764i$ \\
$ 2$& $ 1.1938+1.9457i$ & $ 1.7483+1.8697i$ & $ 2.2915+1.8328i$ \\
$ 3$& $ 0.9944+2.8990i$ & $ 1.5380+2.7393i$ & $ 2.0991+2.6477i$ \\
$ 4$& $ 0.8460+3.9147i$ & $ 1.3365+3.6931i$ & $ 1.8852+3.5353i$ \\
$ 5$& $ 0.7436+4.9483i$ & $ 1.1737+4.6991i$ & $ 1.6828+4.4894i$ \\
$ 6$& $ 0.6711+5.9832i$ & $ 1.0503+5.7271i$ & $ 1.5112+5.4882i$ \\
$ 7$& $ 0.6174+7.0150i$ & $ 0.9565+6.7613i$ & $ 1.3730+6.5100i$ \\
$ 8$& $ 0.5760+8.0431i$ & $ 0.8836+7.7954i$ & $ 1.2627+7.5414i$ \\
$ 9$& $ 0.5429+9.0678i$ & $ 0.8254+8.8275i$ & $ 1.1736+8.5755i$ \\
$10$& $ 0.5158+10.090i$ & $ 0.7778+9.8568i$ & $ 1.1006+9.6093i$ \\
$11$& $ 0.4931+11.109i$ & $ 0.7380+10.883i$ & $ 1.0397+10.641i$ \\
$12$& $ 0.4738+12.126i$ & $ 0.7042+11.908i$ & $ 0.9880+11.671i$ \\
\end{tabular}
%\end{ruledtabular}
\end{table}

%\newpage

%Figures

\begin{figure}[ht]
\centering
\includegraphics[height=8cm,clip]{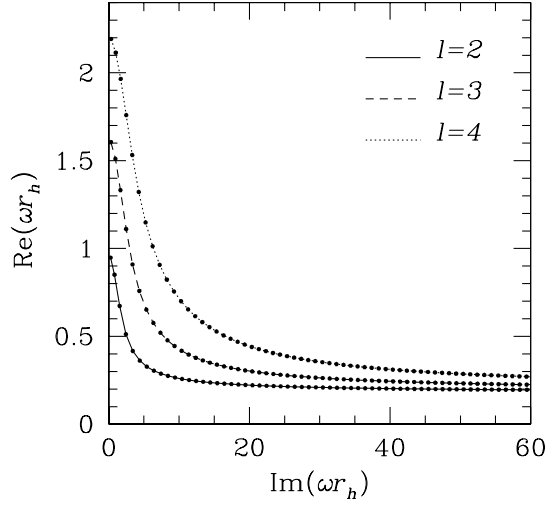}
\caption{Real parts of the non-dimensional QNM frequencies, $\omega r_h$,
given as a function of the imaginary parts of the frequencies for the 
gravitational scalar-gravitational perturbations associated with $l=2$, $l=3$, and 
$l=4$. First sixty QNMs are displayed. The solid circles are used to indicate 
the QN frequencies.}
\end{figure}
\begin{figure}[ht]
\centering
\includegraphics[height=8cm,clip]{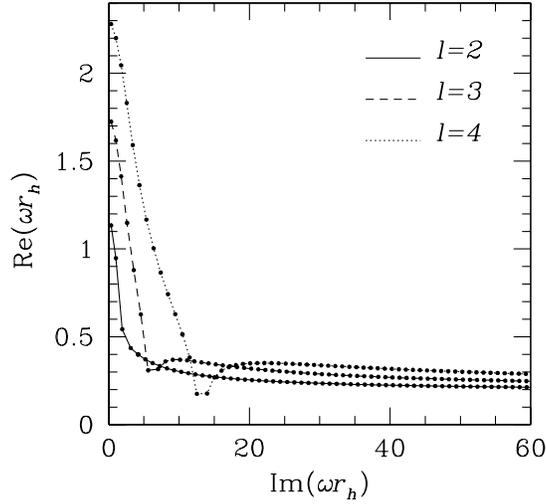}
\caption{Same as Fig. 1 but for vector-gravitational perturbations.}  
\end{figure}
\begin{figure}[ht]
\centering
\includegraphics[height=8cm,clip]{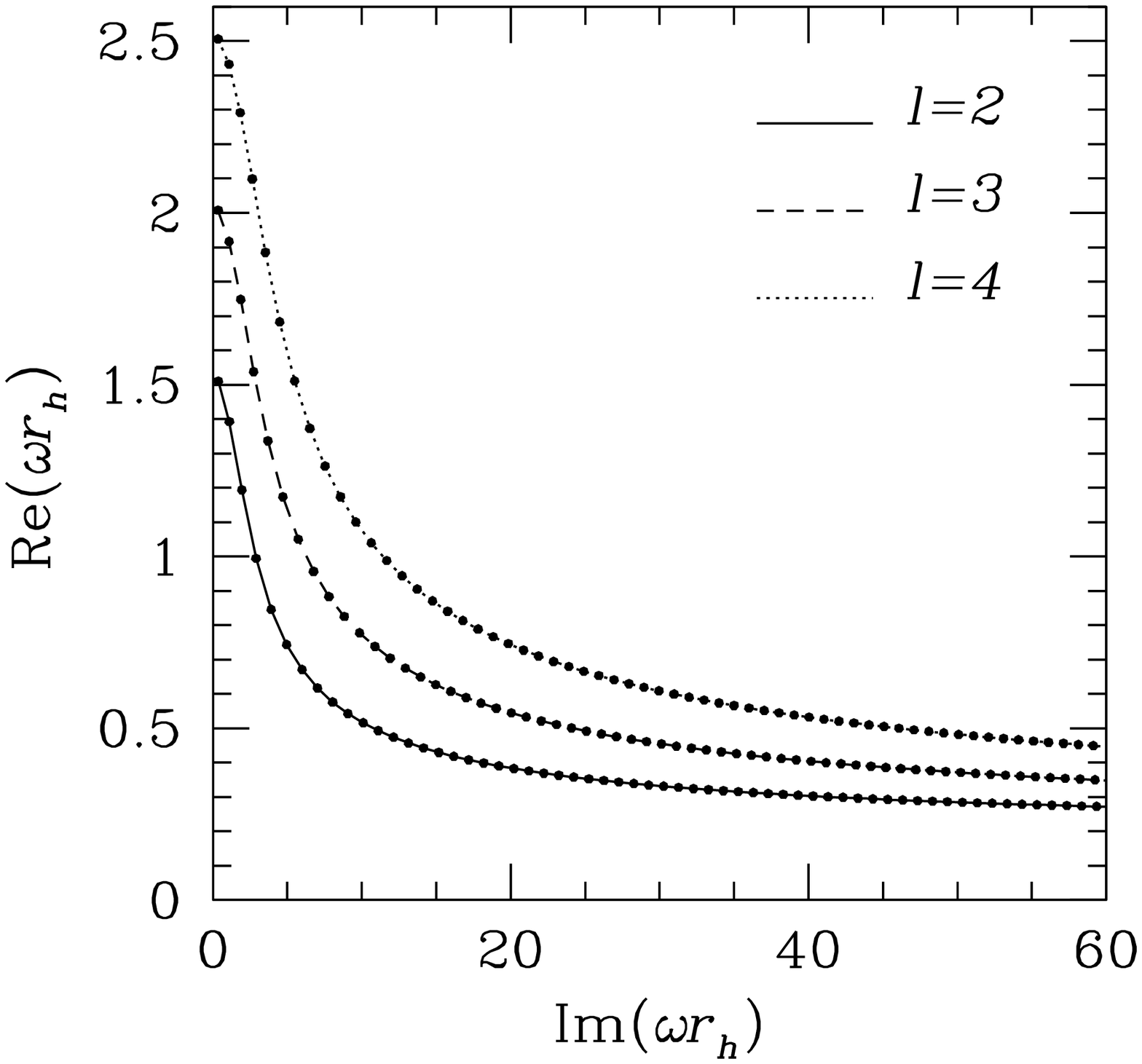}
\caption{Same as Fig. 2 but for the tensor-gravitational perturbations.} 
\end{figure}

In order to confirm that our approach to obtain the QNMs for the 
scalar-gravitational perturbations works well, we first calculate the 
QNMs in the 4D case, namely the QNMs of Zerilli equation. For the Zerilli 
equation, eigenfunction is expanded around the horizon as 
\begin{equation}
\Phi=e^{-i\omega z^{-1}}(z-z_1)^{i\omega z_1^{-1}}z^{i\omega z_1^{-1}}
\sum_{k=0}^\infty a_k\left({z-z_1\over -z_1}\right)^k\,, 
\end{equation}
and the corresponding recurrence relation becomes a five-term relation. 
The numerical result obtained are in good agreement with the QN frequencies 
of the Regge-Wheeler equations, which are exactly the same as those of 
the Zerilli equation, showing that our numerical approach works quite well.

We calculate the QNMs associated with $l=2,3,4$ for the
scalar-gravitational, vector-gravitational, and tensor-gravitational
perturbations in the five dimensional Schwarzschild black with
Nollert's numerical approach \cite{No93}.  In order to discuss the
asymptotic behavior of the QNMs in the limit of large imaginary
frequencies, the modes are obtained up to relatively higher-order modes,
associated with the mode number of $n\sim 300$. Note that the results
for the scalar-gravitational perturbations are newly obtained results,
while the results for the vector-gravitational and
tensor-gravitational perturbations have been already calculated in
\cite{CLY03}.  In the present calculations we have found no
unstable mode whose frequency has a negative imaginary part. This is
consistent with the results by Kodama and Ishibashi \cite{KI03b}, who
proved all the higher-D Schwarzschild black holes to be stable
against small non-radial disturbance.

In Tables 1 through 3, the oscillation frequencies of the thirteen
low-lying QNMs associated with $l=2,3,4$ for the scalar-gravitational,
vector-gravitational, and tensor-gravitational perturbations are,
respectively, tabulated. In the tables, the frequencies are shown in
the units of the inverse of the horizon radius of the black hole
because the horizon radius gives us one of the most natural units for
the low-order QN frequencies.  Note that another convenient
frequency unit, suitable to the asymptotic behavior of highly damped QNMs, 
is the Hawking temperature of the black hole, given by
$T_H=(n-1)/4\pi r_h$ (see, e.g.,
\cite{MN03,Bi03,CLY03}). 

In order to check our numerical code, first
of all, let us compare our results with those obtained by Konoplya
\cite{Ko03b}, who calculated fundamental QNMs in the higher-D
Schwarzschild black hole with the sixth-order JWKB
approximations. Konoplya's frequency of the QNMs associated with $l=3$
for the scalar-gravitational perturbations in the five-dimensional case is
given by $\omega r_h=1.6044+0.3127i$ in the units we use, which is
the only available data for the comparison. It is then found that our
result is in good agreement with that of Konoplya, and that the
relative difference between two frequencies is less than 1\%.

In the 4D Schwarzschild spacetime, there are two
important properties of the QN frequencies: (i) Two types of
gravitational perturbations, namely axial and polar perturbations,
associated with the same angular eigenvalue $l$ possess exactly the
same set of QN frequencies; (ii) There is a purely imaginary
frequency. Interestingly, these two properties are
attributed to the existence of a special relation between effective
potentials of corresponding wave equations for scalar-gravitational 
and vector-gravitational
perturbations. If we write two potentials for scalar-gravitational 
and vector-gravitational
perturbations as $V_+$ and $V_-$, respectively, the existence of this
special relation is equivalent to the existence of a function $F$
satisfying
\begin{equation}
V_{\pm}=\pm f\,{dF\over dr}+F^2+c F\,, 
\label{super}
\end{equation}
where $c$ is some constant \cite{Ch83}. For the higher-D case,
Kodama and Ishibashi found that the function $F$ satisfying equation
(\ref{super}) for scalar- and vector-gravitational perturbations does not exist
in the $D>4$ spacetime \cite{KI03a}. In the five dimensional case,
due to the conclusion of Kodama and Ishibashi, it is expected that the 
three decoupled types of gravitational perturbations have different
sets of the QN frequencies, and that there is no purely
imaginary frequency of the QNMs. Our numerical results are consistent
with these predictions.  On Tables 1 through 3, it is observed that
the three types of gravitational perturbations do not have the same set of
QN frequencies even when the modes are associated with
the same angular quantum number $l$. We also found no purely imaginary
frequency of the QNMs.

In Figs. 1 through 3, we also display the real parts of the
frequencies, ${\rm Re}(\omega r_h)$, versus the the imaginary parts of
the frequencies, ${\rm Im}(\omega r_h)$, for the oscillation
frequencies of the lower-lying QNMs associated with $l=2,3,4$ for the
scalar-gravitational, vector-gravitational, and tensor-gravitational
perturbations, respectively. In these figures, the first sixty QNM
frequencies are exhibited.  In Figs. 1-3, we can observe that the real
parts of the QNMs frequencies are monotonically decreasing functions
of the imaginary parts of the frequencies except for the cases of the
vector-gravitational perturbations associated with $l=3$ and $l=4$. The
behavior of the QNMs of the vector-gravitational perturbations associated with
$l=3$ and $l=4$ are similar to those of the gravitational
perturbations in the 4D Schwarzschild black hole: the
real parts of frequencies first decrease to a minimum value,
increase to the local maximum, and then approach the asymptotic
constant values, as the imaginary parts of the frequencies are
increased.

Finally, let us discuss the asymptotic behavior of the QNMs in the
limit of highly damped modes. In the Schwarzschild black hole in the
$D=n+2$ dimensional spacetime, the asymptotic QN frequencies in the
limit of highly damped modes are given by
\begin{eqnarray}
\omega T_H^{-1}=\log{3}+2\pi i (n+{1\over2})\,,
\label{asymptotic}
\end{eqnarray}
regardless of the spacetime dimensions and the angular quantum number
$l$ \cite{MN03,Bi03,CLY03}.  On the other hand, higher-order
corrections of the frequency (\ref{asymptotic}) depend both on the
spacetime dimensions and the angular quantum number $l$
\cite{CLY03}. For the vector-gravitational and tensor-gravitational
perturbations, the asymptotic behavior were already investigated in
detail by Cardoso, Lemos and Yoshida \cite{CLY03}. Thus, we only
consider the case of the scalar-gravitational perturbations, here.  As
shown in equation (\ref{asymptotic}), we have ${\rm Re}(\omega
r_h)\rightarrow (\log{3})/2\pi$ as ${\rm Im}(\omega r_h)\rightarrow
\infty$ for the 5D case. Fig. 1 shows that the real parts of the QN
frequencies approach to the asymptotic constant value $(\log{3})/2\pi$
as the imaginary parts of the frequencies increase, and that our
numerical results are consistent with the analytical prediction of the
asymptotic behavior in the limit of large imaginary
frequency. Unfortunately, we cannot however obtain the QNMs associated
with sufficiently large mode number with our numerical code. The
reason of this is unclear, but we may guess that this is because the
recurrence relation (\ref{recu}) for obtaining the QN frequencies
becomes eight-term relation in the case of the scalar-gravitational
perturbations. We therefore have failed to extract the detailed
information of the asymptotic behavior of the QNM in the
scalar-gravitational perturbations.

\section{Conclusion}

We have calculated the QNMs for the three types of gravitational
perturbations of the 5D Schwarzschild black hole through a continued
fraction method. In order to examine the QNMs, we made use of
Schr\"odinger-type wave equations for determining the dynamics of the
gravitational perturbations \cite{KI03a}.  To apply the continued
fraction method, we expanded the eigenfunctions around the black hole
horizon in terms of a Fr\"obenius series. It was found that
scalar-gravitational perturbations obey an eight-term recurrence
relation, and vector-gravitational and tensor-gravitational
perturbations obey a four-term recurrence relation.  For all the types
of perturbations, the QNMs associated with $l=2$, $l=3$, and $l=4$
were calculated in the present study.  Our numerical results are
summarized as follows; (i) The three types of gravitational
perturbations have different QNM frequencies; (ii) There is no purely
imaginary frequency mode (the above two results are consistent with
Kodama and Ishibashi's results that there is no function satisfying
equation (\ref{super})); (iii) The QNMs belonging to the 
three types of gravitational
perturbations have the same asymptotic behavior in the
limit of the large imaginary frequencies, given by $\omega
T_H^{-1}\rightarrow\log{3}+2\pi i (n+1/2)$ as $n\rightarrow\infty$; This
asymptotic behavior was already expected \cite{MN03,Bi03}.

For the black hole dynamics, the lower-order QNMs play an important
role after the initial non-linear effects have become
insignificant. In the 4D Schwarzschild black hole, one can guess which
frequency will be excited in the ringing phase since the axial and
polar QNMs have the same frequencies. On the other hand, there are
three different sets of QNM frequencies for the Schwarzschild black
hole in $D\ge 5$ dimensions. Our results show that the fundamental QNM
for the scalar-gravitational perturbations have the smallest frequency
and the longest damping-time. This means that the QNMs for the
scalar-gravitational perturbations can live longer than other types of
perturbations if all the types of perturbations are excited with the
same initial amplitudes. Probably, in general situations things are
not so simple.  What type of perturbations will be excited will be
strongly dependent on the situation. In order to expect what frequency
will be excited in the ringing phase, one has to model the source
terms of the master equations. So far, only the case where a test 
particle radially falls into a black hole has been considered \cite{BCG03}. 
In this case, due to the symmetry of the motion, only the 
scalar-gravitational perturbations are excited. The general case 
remains as a work to be solved.

\vskip 2mm

\section*{Acknowledgements}

This work was partially funded by Funda\c c\~ao para a Ci\^encia e
Tecnologia (FCT) -- Portugal through project CERN/FIS/43797/2001. SY
acknowledges financial support from FCT through project SAPIENS
36280/99.  VC also acknowledge financial support from FCT through
PRAXIS XXI programme.  JPSL thanks Observat\'orio Nacional do Rio de
Janeiro for hospitality.

%\newpage 
%


\begin{thebibliography}{99}
%
\bibitem{vish} C. V. Vishveshwara, Nature {\bf 227}, 936 (1970).
%
\bibitem{kokkotas99} K. D. Kokkotas and B. G. Schmidt, Living Rev. Relativity 
{\bf 2}, 2 (1999).
%
\bibitem{nollert99} H. P. Nollert, Class. Quant. Grav. {\bf 16}, R159 (1999).
%
\bibitem{hod} J. Bekenstein,
Lett. Nuovo Cimento {\bf 11}, 467 (1974);
J. Bekenstein and V. F. Mukhanov, 
Phys. Lett. B {\bf 360}, 7 (1995);
S. Hod, 
Phys. Rev. Lett. {\bf 81}, 4293 (1998);
O. Dreyer,
Phys. Rev. Lett. {\bf 90}, 081301 (2003);
J. Oppenheim, gr-qc/0307089;
A. Corichi, 
Phys.Rev. D {\bf 67}, 087502 (2003);
 Y. Ling, H. Zhang, gr-qc/0309018;
%
\bibitem{CLY03} V. Cardoso, J. P. S. Lemos, and S. Yoshida, gr-qc/0309112.
%
\bibitem{All} V. Cardoso and J. P. S. Lemos,
Phys. Rev. D {\bf 67}, 084020 (2003);
C. Molina,
Phys. Rev. D {\bf 68}, 064007 (2003);
A. M. van den Brink, 
Phys. Rev. D {\bf 68}, 047501 (2003);
V. Cardoso, R. Konoplya, J. P. S. Lemos,
Phys. Rev. D {\bf 68}, 044024 (2003);
E. Berti, K. D. Kokkotas, E. Papantonopoulos, Phys. Rev. D {\bf 68}, 
 064020 (2003);
E. Berti, V. Cardoso, K. D. Kokkotas, H. Onozawa
hep-th/0307013; E. Berti and K. D. Kokkotas, 
Phys. Rev. D {\bf 68}, 044027 (2003);
D. Birmingham, S. Carlip and Y. Chen,
Class. Quant. Grav. {\bf 20}, L239 (2003);
D. Birmingham,
Nucl. Phys. {\bf B 671},  67 (2003);
N. Andersson, C. J. Howls,
gr-qc/0307020;
S. Yoshida and T. Futamase,
gr-qc/0308077;
S. Musiri and G. Siopsis, hep-th/0308196;
hep-th/0308168; hep-th/0309227;
D. Birmingham, S. Carlip, hep-th/0311090;
K. H. C. Castello-Branco and E. Abdalla, gr-qc/0309090;
A. J. M. Medved, D. Martin and M. Visser, gr-qc/0310009; gr-qc/0310097;
T. R. Choudhury and T. Padmanabhan, gr-qc/0311064;
T.Padmanabhan, gr-qc/0311036;
M. R. Setare, hep-th/0311221.
%
\bibitem{hamed} N. Arkani-Hamed, S. Dimopoulos and G. Dvali,
Phys. Lett. {\bf B 429}, 263 (1998); Phys. Rev. D {\bf 59}, 086004 (1999);
I. Antoniadis, N. Arkani-Hamed, S. Dimopoulos and G. Dvali,
Phys. Lett. {\bf B 436}, 257 (1998). 
%
\bibitem{CDL03} V. Cardoso, O. J. Dias and J. P. S. Lemos, Phys. Rev. D 
{\bf 67}, 064026 (2003). 
%
\bibitem{IUM03} D. Ida, Y. Uchida, and Y. Morisawa, Phys. Rev. D {\bf 67}, 
084019 (2003).
%
\bibitem{Ko03a} R. A. Konoplya, Phys. Rev. D {\bf 68}, 024018 (2003). 
%
\bibitem{MN03} L. Motl and A. Neitzke, Adv. Theor. Math. Phys. {\bf 7}, 
 307 (2003).
%
\bibitem{KI03a} H. Kodama and A. Ishibashi, Prog. Theor. Phys. {\bf 110}, 
 701 (2003). 
%
\bibitem{Ko03b} R. A. Konoplya, hep-th/0309030.
%
\bibitem{BCG03} E. Berti, M. Cavagli\`a, and L. Gualtieri, hep-th/0309203. 
%
\bibitem{Bi03} D. Birmingham, Phys. Lett. {\bf B 569}, 199 (2003).
%
\bibitem{RW57} T. Regge and J. A. Wheeler, Phys. Rev. {\bf 108}, 1063 (1957).
%
\bibitem{Ze70} F. J. Zerilli, Phys. Rev. Lett. {\bf 24}, 737 (1970).
%
\bibitem{No93} H.-P. Nollert, Phys. Rev. D {\bf 47}, 5253 (1993). 
%
\bibitem{Le85} E. W. Leaver, Proc. R. Soc. London {\bf A402}, 285 (1985).
%
\bibitem{Le90} E. W. Leaver, Phys. Rev. D {\bf 41}, 2986 (1990). 
%
\bibitem{Gu67} W. Gautschi, SIAM Rev. {\bf 9}, 24 (1967). 
%
\bibitem{KI03b} H. Kodama and A. Ishibashi, gr-qc/0305185.
%
\bibitem{Ch83} S. Chandrasekhar, {\it The mathematical theory of black 
holes} (Clarendon Press, Oxford, 1983).  

\end{thebibliography}
\end{document}